\documentclass{PoS}

\title{THEORETICAL REACTION RATES OF THE $^{12}$C($\alpha$,$\gamma$)$^{16}$O REACTION FROM THE POTENTIAL MODEL}

\ShortTitle{THEORETICAL REACTION RATES OF THE $^{12}$C($\alpha$,$\gamma$)$^{16}$O REACTION FROM THE POTENTIAL MODEL}

\author{\speaker{Masahiko Katsuma}%
        \\
        Advanced Mathematical Institute, Osaka City University\\
        E-mail: \email{mkatsuma@sci.osaka-cu.ac.jp}}

\abstract{
  The radiative capture cross sections of $^{12}$C($\alpha$,$\gamma$)$^{16}$O and derived reaction rates are calculated from the direct capture potential model.
  The resulting $S$-factor at low energies is found to be dominated by $E$2 transition to the $^{16}$O ground state.
  The $E$1 and $E$2 $S$-factors at $E_{c.m.}=0.3$ MeV are $S_{E1}\approx3$ keV~b and $S_{E2}=150^{+41}_{-17}$ keV~b, respectively.
  The sum of the cascade transition through the excited state of $^{16}$O is $S_{\rm casc}= 18\pm4.5$ keV~b.
  The derived reaction rates at low temperatures seem to be concordant with those from the previous evaluation.
  For astrophysical applications, our reaction rates below $T_9=3$ are provided in an analytic expression.
}

\FullConference{XIII Nuclei in the Cosmos\\
		 7-11 July, 2014\\
		 Debrecen, Hungary}

\begin{document}

\section{Introduction}
\label{sec-1}
  The $^{12}$C($\alpha$,$\gamma$)$^{16}$O reaction has been considered to be one of the most important nuclear reactions in the nucleosynthesis of elements in the universe \cite{Rol88}.
  However, due to the Coulomb barrier, the cross section at the center-of-mass energy $E_{c.m.} = 0.3$ MeV corresponding to helium burning temperatures is too small to measure using present laboratory technologies.
  In this circumstance, the low-energy cross section is extrapolated from the available experimental data with theoretical guides, and it is converted into the nuclear reaction rates for astrophysical applications.

  The representative studies of the $^{12}$C($\alpha,\gamma)^{16}$O reaction rates have been provided by KU02 \cite{Kun02} and BU96 \cite{Buc96}, based on the $R$-matrix analyses.
  For the extrapolated astrophysical $S$-factor at $E_{\rm c.m.}= 0.3$ MeV, which is used instead of the cross section, KU02 estimates $S_{E1}=76\pm20$ keV~b, $S_{E2}=85\pm30$ keV~b, and $S_{\rm casc}=4\pm4$ keV~b for the $E$1, $E$2, and cascade transitions.
  BU96 predicts $S_{E1}=79\pm21$ keV~b, $S_{E2}=70\pm70$ keV~b, and $S_{\rm casc}=16\pm16$ keV~b.
  The derived reaction rates have been reported to give a different temperature dependence from that in the pioneering Fowler compilation of the reaction rates (CF88 \cite{CF88}).
  The so-called NACRE compilation \cite{NACRE} following CF88 is the milestone of the compilation work, providing the evaluated rates to the astrophysical community.

  In the present study, we illustrate the calculated results of the $S$-factor and $\gamma$-ray angular distribution, and we compare the derived reaction rates (KA12) \cite{Kat12} with the previous studies (CF88, NACRE, KU02, BU96).
  We use the potential model \cite{Kat12,Kat10}.
  This is the one of direct nuclear reaction models, which describe the reaction process, in which a few degrees of freedom of motion are activated.
  Finally, we show an analytic expression of our reaction rates.

\section{Potential models}
\label{sec-2}

  We first describe a brief outline of our direct capture potential model \cite{Kat12,Kat10}.
  The radiative capture cross section $\sigma$ is calculated from $\sigma \propto |\langle \varphi_f | M^E_\lambda| \varphi_i \rangle |^2$, where $\varphi_i$, $\varphi_f$, and $M^E_\lambda$ are the scattering wave, the bound state, and electromagnetic operator, respectively.
  $\lambda$ denotes multipolarity of the transition.
  The input parameters of the model come from the potential used to generate $\varphi_i$ and $\varphi_f$.
  The effective charge in $M^E_\lambda$ is also used as an adjustable parameter to minimize the difference between the theoretical values and the experimental cross section data.
  The astrophysical $S$-factor is defined as $S_{E\lambda}=E_{c.m.} \exp(2\pi\eta) \sigma$, where $\eta=12e^2/\hbar v$.
  $v$ is the relative velocity.

\begin{figure}[t]
    \centering
    \includegraphics[width=0.38\linewidth,clip]{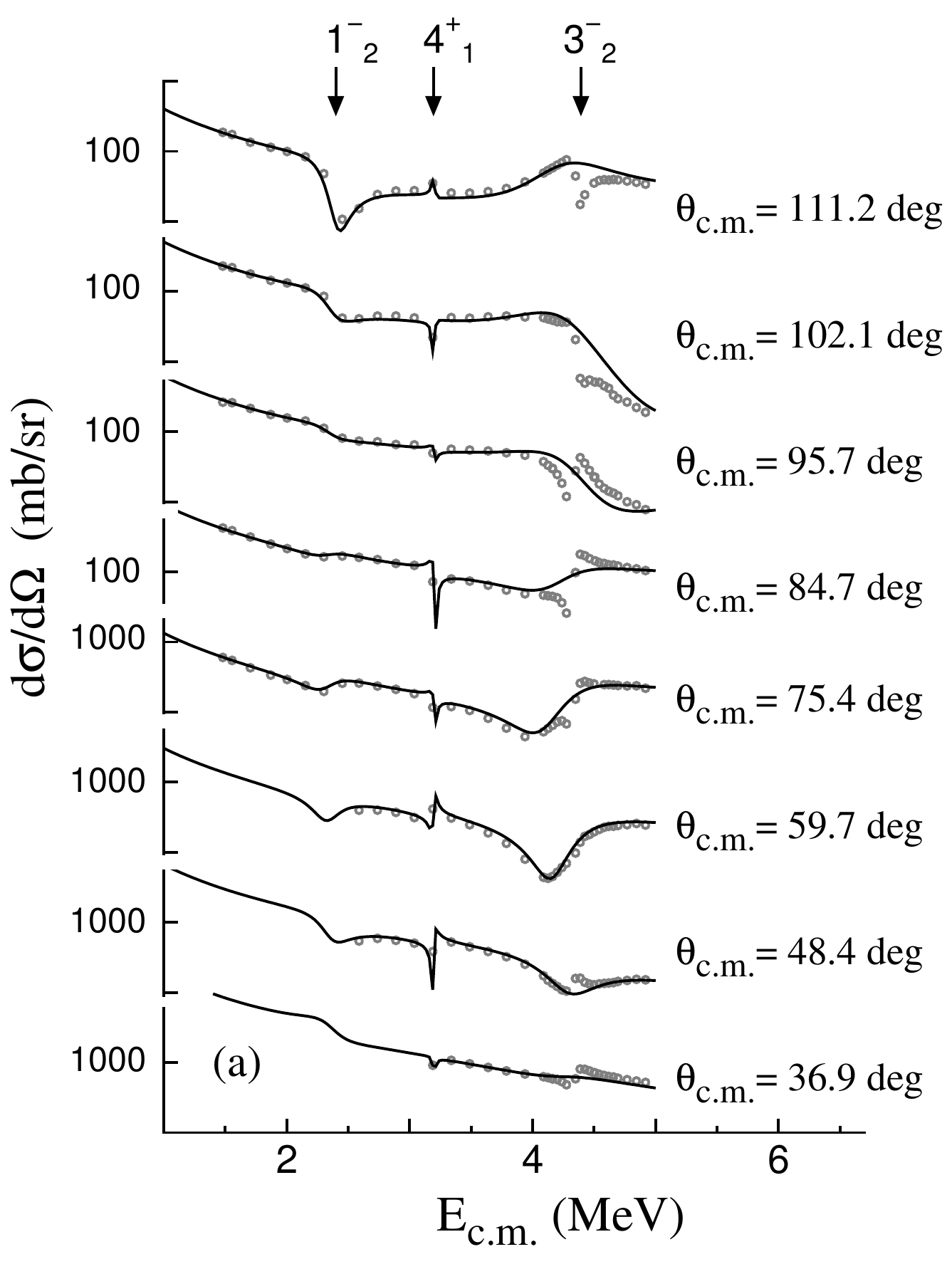}
    \hspace{3mm}
    \includegraphics[width=0.38\linewidth,clip]{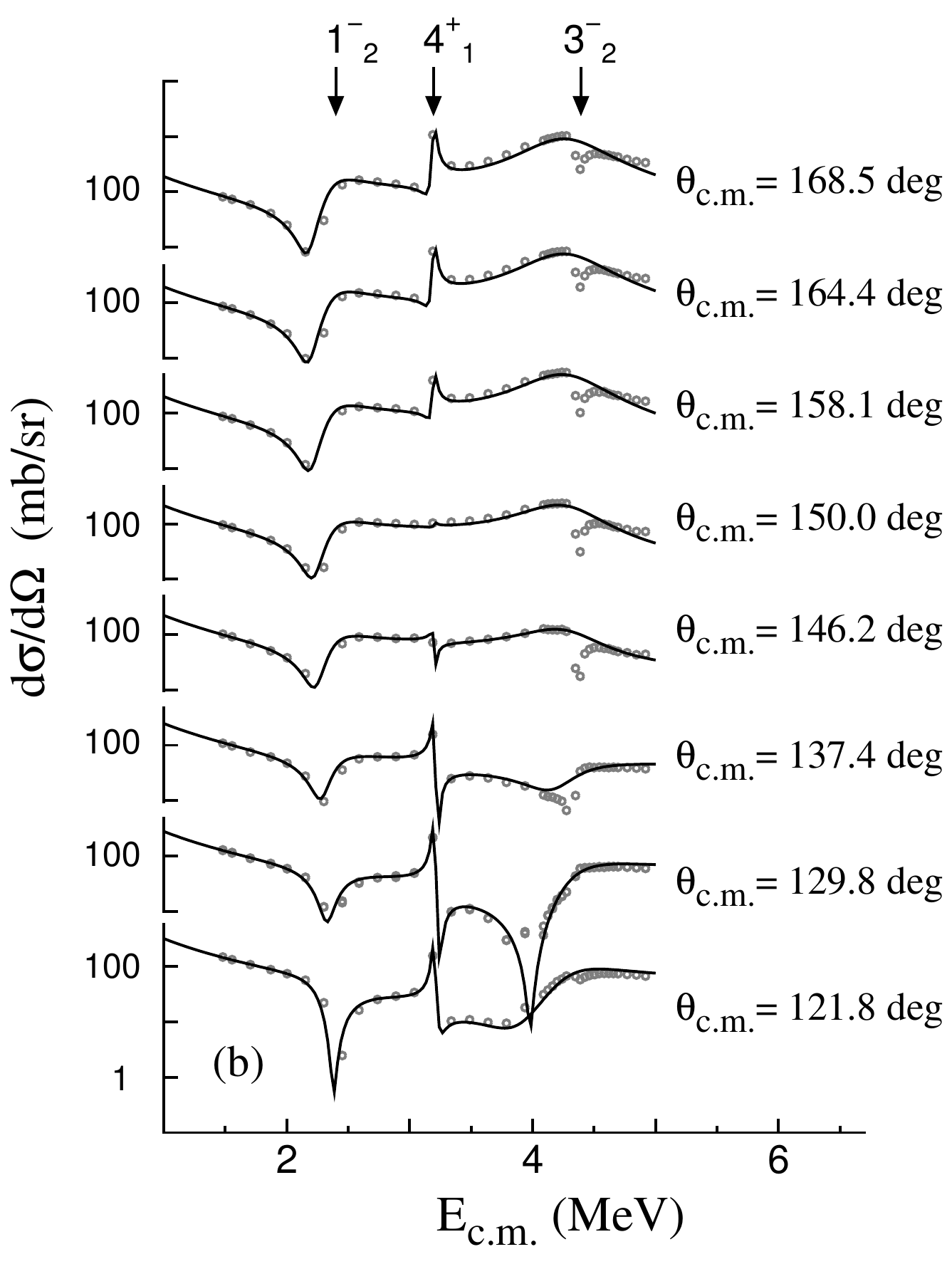}
    \caption{\label{fig:0}       
      The excitation functions of $\alpha$+$^{12}$C elastic scattering below $E_{c.m}=5$ MeV \cite{Kat10}.
      The solid curves are the calculated results with potential scattering.
      The arrows indicate the resonance energies of the states in the rotational bands.
      The experimental data are taken from \cite{Pla87}.
    }
\end{figure}

  The $\varphi_i$ is generated from the potential that describes phase shifts and differential cross sections for $\alpha$+$^{12}$C elastic scattering \cite{Kat12,Kat10,Pla87,Tis09}.
  The internuclear potential should be chosen in the appropriate strength.
  To obtain the correct strength, we refer to nuclear rainbow scattering at high energies \cite{Ing94,Bra97,Sat83}.
  The potential used here resembles the double-folding potential with M3Y interaction \cite{Ber77}, and it produces the $\alpha$+$^{12}$C elastic cross sections shown in Fig.~\ref{fig:0}.
  For the bound states $\varphi_f$, we classify the states into two groups.
  One is the states well described by $\alpha$-cluster models. (0$^+_2$: $E_x= 6.05$ MeV, and 2$^+_1$: $E_x=6.92$ MeV.)
  $E_x$ denotes the excitation energy of $^{16}$O.
  To make the $\alpha$-cluster states, we use the potential for the $\alpha$+$^{12}$C rotational bands \cite{Kat10}.
  Another is the shell model states. (0$^+_1$: ground state, 3$^-_1$: $E_x= 6.13$ MeV, and 1$^-_1$: $E_x=7.12$ MeV.)
  To make these, we use the so-called separation energy method, in which the potential is adjusted so as to reproduce the experimental separation energy.
  This is the same as a textbook method in direct reaction models, e.g. distorted wave Born approximation (DWBA) \cite{Sat83}.
  It gives appropriate wavefunctions in the peripheral region that nuclear reactions are sensitive to.
  The large difference of the potential between the bound state and scattering state may indicate that the wavefunction is not appropriate in the assumed configuration.
  However, it is good for a description of direct reactions as a doorway to making fused nuclei from scattering states.

  The Maxwellian-averaged reaction rates $N_A \langle\sigma v\rangle$ \cite{Rol88,NACRE} are calculated from the theoretical cross section by the potential model.

\section{Results}
\label{sec-3}

  Before comparing the reaction rates with CF88, NACRE, KU02, and BU96, we show the theoretical calculation of the $S$-factor and $\gamma$-ray angular distribution from the potential model.

\begin{figure}[t]
  \begin{minipage}[b]{0.48\linewidth}
    \centering
    \includegraphics[width=0.81\linewidth,clip]{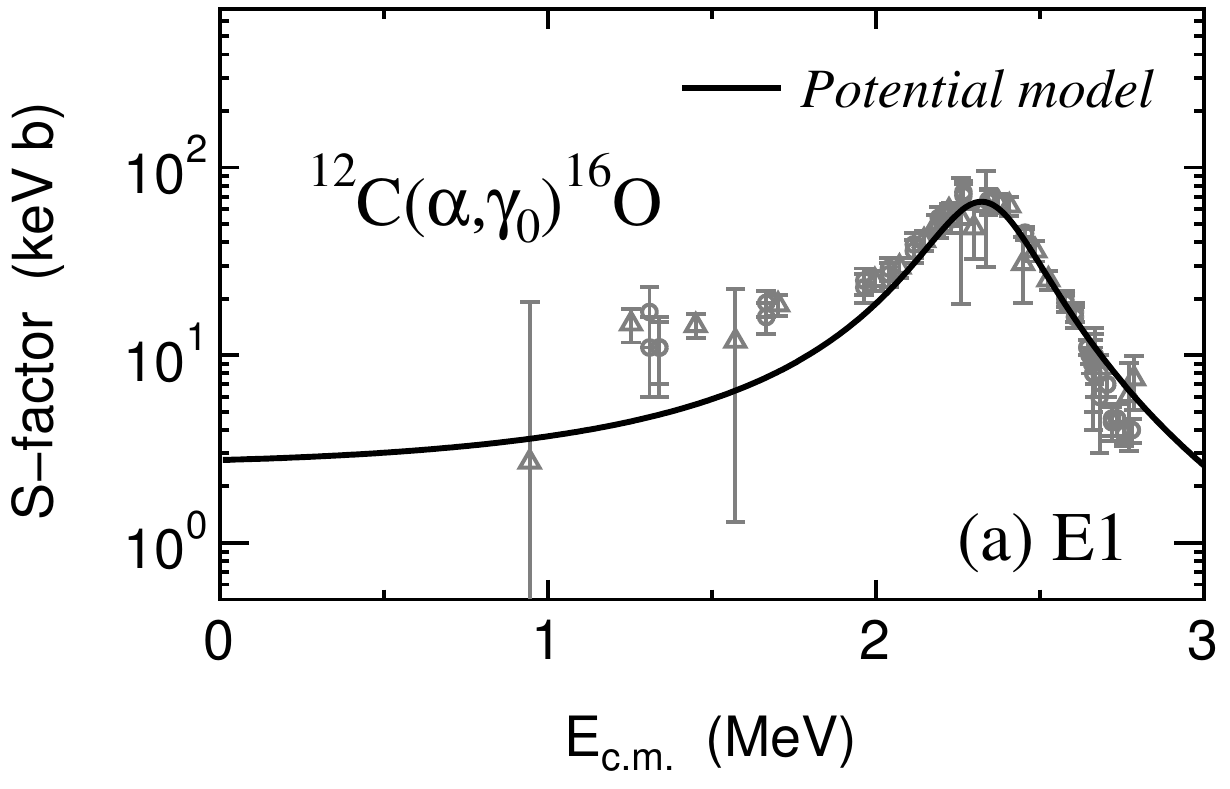}
    \includegraphics[width=0.81\linewidth,clip]{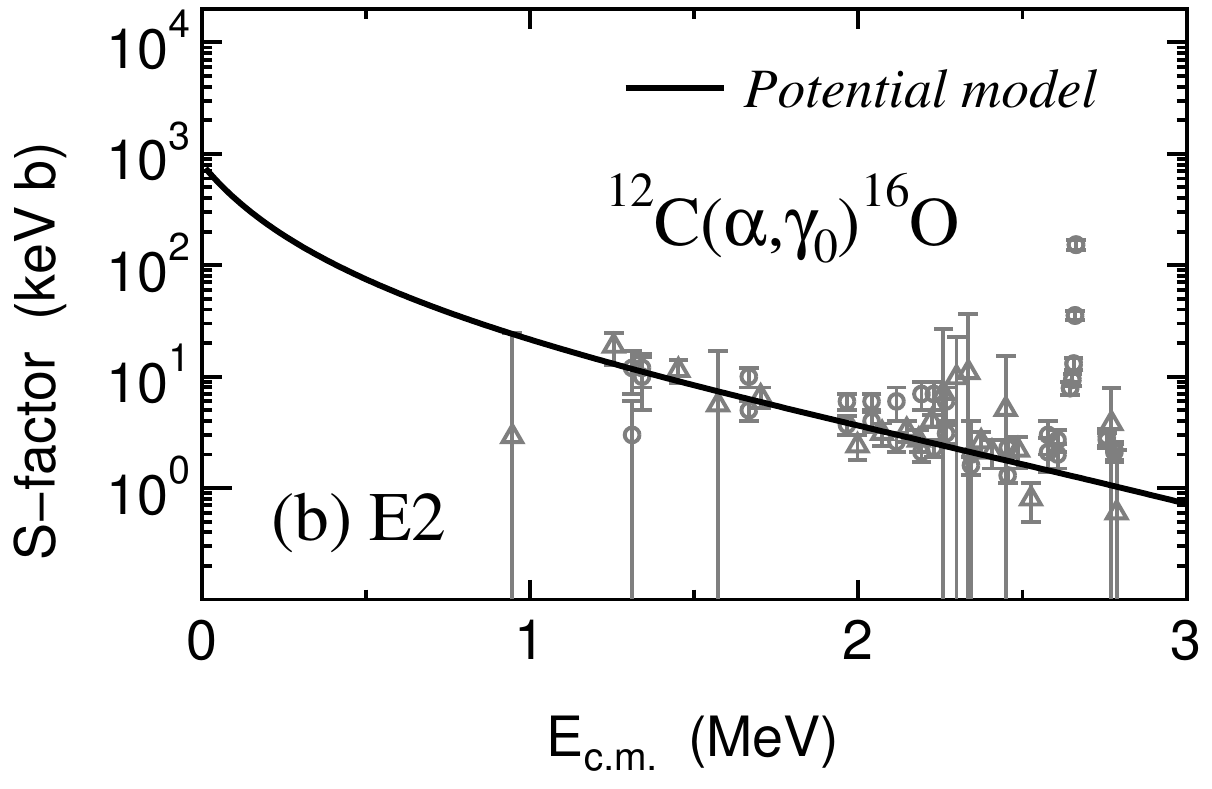}
    \caption{\label{fig:1}       
      The astrophysical $S$-factor of the $^{12}$C($\alpha$,$\gamma_0$)$^{16}$O reaction: (a) $E$1 and (b) $E$2.
      The solid curves are the results obtained from the potential model.
      The experimental data are taken from \cite{Kun02,Ass06}.
    }
  \end{minipage}
  \hfill
  \begin{minipage}[b]{0.48\linewidth}
    \centering
    \includegraphics[width=0.66\linewidth,clip]{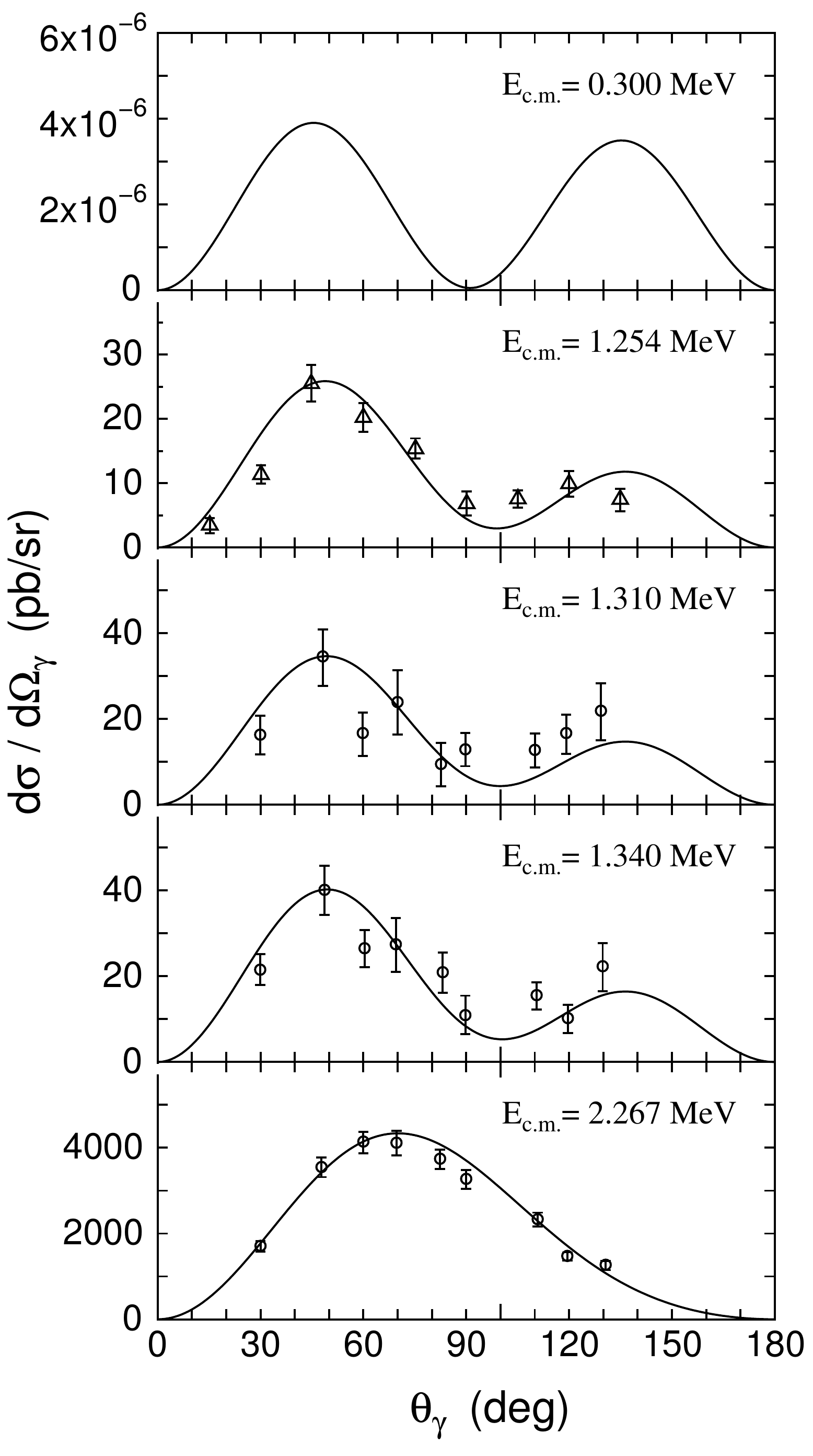}
    \caption{\label{fig:2}       
      The $\gamma$-ray angular distribution of the $^{12}$C($\alpha$,$\gamma_0$)$^{16}$O reaction \cite{Kat12}.
      The experimental data are taken from \cite{Kun02,Ass06}.
    }
  \end{minipage}
\end{figure}

  The calculated $S$-factor for $E$1 and $E$2 transitions of the $^{12}$C($\alpha$,$\gamma_0$)$^{16}$O reaction is compared with the experimental data \cite{Kun02,Ass06} in Fig.~\ref{fig:1}.
  From this figure, the $E$2 transition is found to be enhanced at low energies.
  The $\gamma$-ray angular distribution of the $^{12}$C($\alpha$,$\gamma_0$)$^{16}$O reaction is shown in Fig.~\ref{fig:2}.
  The potential model (solid curves) appears to reproduce the trend of the peak and valley of the $\gamma$-ray angular distribution.
  At $E_{c.m.}= 2.267$ MeV, the $\gamma$-ray angular distribution has a single peak, that means the p-wave dominates the transition.
  At low energies, the d-wave seems to become more important.
  Although the $E$1 component appears to deviate from the experimental data in Fig.~\ref{fig:1}(a), the angular distribution seems to be reproduced by the potential model.
  The uncertainties in the experimental $\gamma$-ray angular distribution have recently been discussed in \cite{Pla12}.
  The extrapolated values of the $S$-factor at $E_{c.m.}=0.3$ MeV are listed in Table~\ref{tab:1}.
  The total $S$-factor from the present work is consistent with KU02, NACRE, BU96.
  In contrast, our result for the $E$1 and $E$2 components is different from the previous studies.
  However, let us recall that the decomposition of $E$1 and $E$2 is performed from the analysis of the $\gamma$-ray angular distribution.

  The subthreshold 2$^+_1$ state and the resonant 1$^-_2$ state at $E_x\approx9.6$ MeV are the members of the $\alpha$+$^{12}$C rotational bands in $^{16}$O \cite{Kat12,Kat10}.
  They appear to have the broad width of electric transition, as shown in Fig.~\ref{fig:1}.
  The narrow resonances do not interfere with the present result so much, and they may be negligible \cite{Kat14}.
  The $R$-matrix may not give a good description of the $^{12}$C($\alpha$,$\gamma_0$)$^{16}$O reaction at the astrophysical energies, because it cannot explain e.g. the weak coupling of the system, the state with a broad width, and large violation of the isospin selection rule \cite{Kat14}.
  The experimental studies of the total cross sections \cite{Sch05} have also been performed, recently.

  The asymptotic normalization constant (ANC) can be obtained phenomenologically,
  $C^2=5.00\times10^{28}$ fm$^{-1}$ for 1$^-_1$, $C^2=2.03\times10^{10}$ fm$^{-1}$ for 2$^+_1$.
  These values are consistent with the previous studies \cite{Spa04,Bru99}.

\begin{table}[t]
  \centering
  \caption{
    Astrophysical $S$-factor at $E_{c.m.}=0.3$ MeV.
    The $S$-factor is listed in keV~b unit.
    The total $S$-factor includes the contribution from the cascade transition through the excited states of $^{16}$O.
  }
  \label{tab:1}
  \begin{tabular}{ccccc}
    \hline
    & This work     & KU02 \cite{Kun02} & NACRE \cite{NACRE} & BU96 \cite{Buc96}\\
    \hline
    $E$1      &  $3$            & $76\pm20$  & $79\pm21$  &  $79\pm21$\\
    $E$2      & $150^{+41}_{-17}$ & $85\pm30$  & $120\pm60$ &  $70\pm70$\\
    Cascade & $18\pm4.5$      &  $4\pm4$   &      -      &  $16\pm16$\\
    Total   & $171^{+46}_{-22}$ & $165\pm54$ & $199\pm81$ & $165\pm107$\\
    \hline
  \end{tabular}
\end{table}

\begin{figure}[t]
  \centering
  \includegraphics[width=0.40\linewidth,clip]{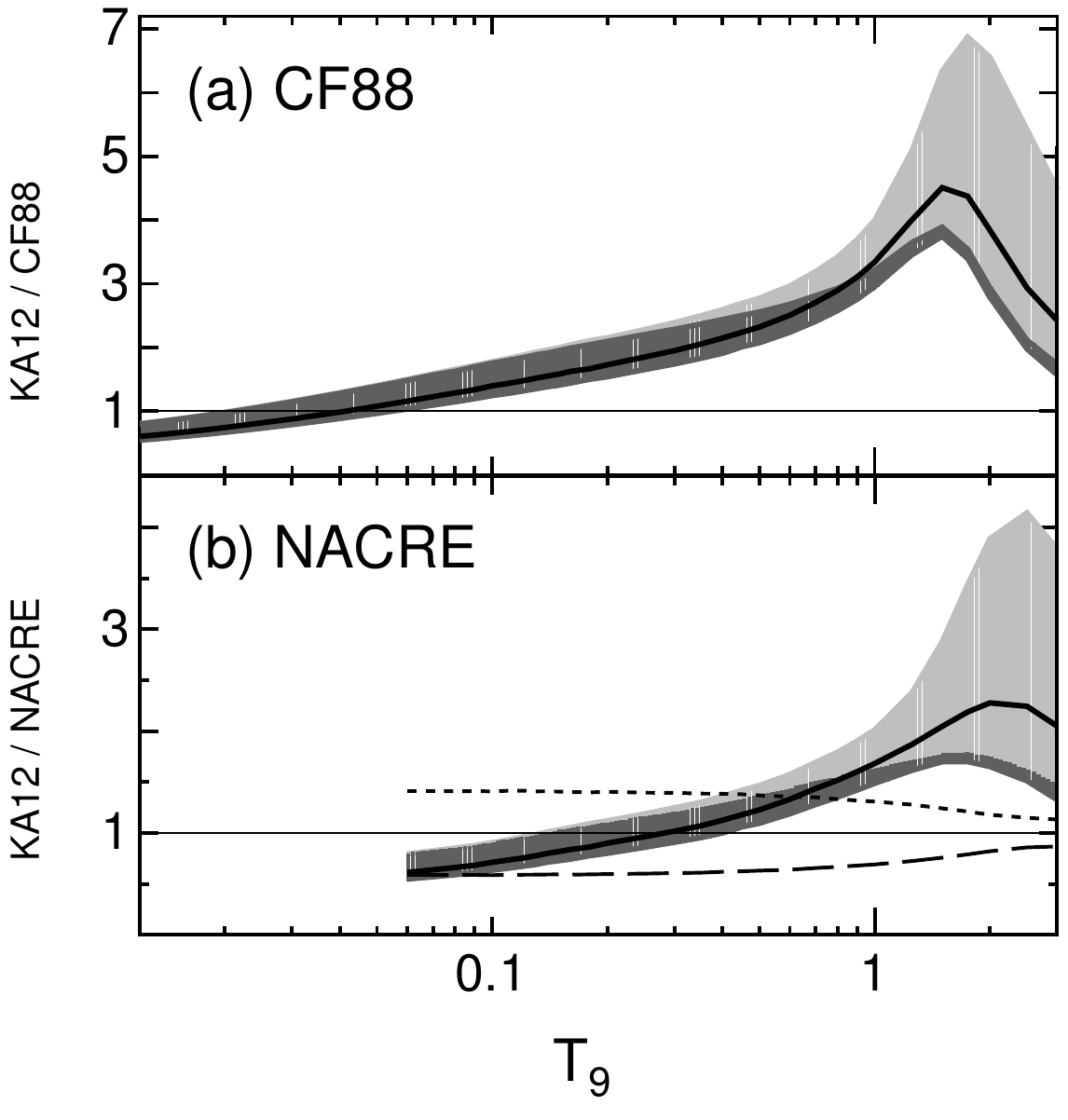}
  \hspace{5mm}
  \includegraphics[width=0.40\linewidth,clip]{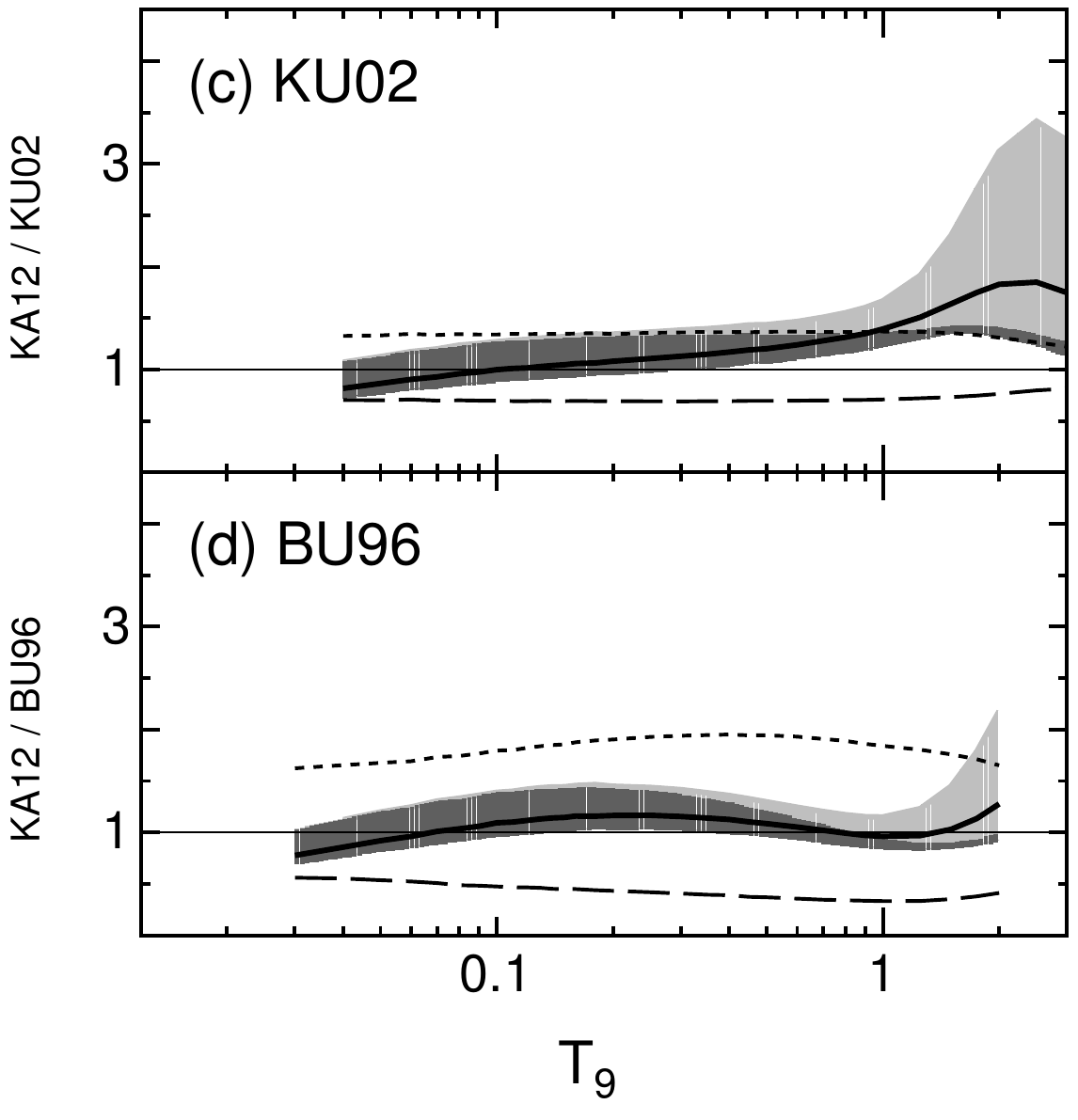}
  \caption{
    Comparison between our reaction rate KA12 and one from the previous studies.
    The reaction rates are expressed in the ratio to (a) CF88, (b) NACRE, (c) KU02, and (d) BU96.
    The solid curves are the recommended reaction rates.
    The shades are uncertainties estimated from the model parameters.
    The dense and thin shades come from the variation of the potential parameters and the strength of the cascade transition.
    The dotted and dashed curves are the upper and lower limits estimated by the respective group.
  }
  \label{fig:3}       
\end{figure}

  In Fig.~\ref{fig:3} the derived reaction rates (KA12) are compared with those from the representative studies.
  The panel (a), (b), (c), and (d) show the ratio of the reaction rates to CF88, NACRE, KU02, and BU96, respectively.
  We find from this figure that KA12 seems to be close to BU96.
  The theoretical uncertainties of the reaction rates are estimated from the variation of the model parameters \cite{Kat12}.
  The dark area comes from the variation of the potential parameters; the thin hatch comes from the strength of the cascade transition of $^{16}$O.

  Our reaction rates below $T_9=3$ (KA12) are approximately given in the simple expression in unit of ${\rm cm}^3{\rm mol}^{-1}{\rm s}^{-1}$,
  \begin{eqnarray}
    N_A \langle\sigma v\rangle 
    &=&
    1.16\!\times\!10^9 \,T_9^{-4/3} 
    \exp\left[-\frac{32.369}{T_9^{1/3}}-\left(\frac{T_9}{3.17}\right)^2\,\right] 
    \Big(\, 1 -0.371 \,T_9 -0.106 \,T_9^2 +0.264 \,T_9^3 \Big),
    \nonumber
  \end{eqnarray}
  where $T_9$ is temperature in units of $10^9$ K.
  The upper and lower limits, $N_A \langle\sigma v\rangle_{\rm H}$ and $N_A \langle\sigma v\rangle_{\rm L}$, are expressed by 
  \begin{eqnarray}
    N_A \langle\sigma v\rangle_{\rm H} 
    &=&     
    1.32\!\times\!10^9 \,T_9^{-4/3} 
    \exp\left[-\frac{32.330}{T_9^{1/3}}-\left(\frac{T_9}{2.80}\right)^2\,\right] 
    \Big(\, 1 -0.0686 \,T_9 -0.845 \,T_9^2 +0.711 \,T_9^3 \Big),
    \nonumber\\
    N_A \langle\sigma v\rangle_{\rm L} 
    &=&     
    1.02\!\times\!10^9 \,T_9^{-4/3} 
    \exp\left[-\frac{32.363}{T_9^{1/3}}-\left(\frac{T_9}{4.04}\right)^2\,\right] 
    \Big(\, 1 -0.422 \,T_9 +0.124 \,T_9^2 +0.0894 \,T_9^3 \Big).
    \nonumber
  \end{eqnarray}
  The tabular form of the reaction rates can be found in \cite{Kat12}.

\section{Summary}
\label{sec-4}

  The astrophysical $S$-factor of $^{12}$C($\alpha$,$\gamma$)$^{16}$O has been investigated with the potential model.
  The derived reaction rates have been compared with those from CF88, NACRE, KU02, and BU96.

  The total $S$-factor obtained from the potential model is concordant with NACRE, KU02, and BU96.
  However, the E2/E1 ratio is different from that of the previous works.
  From the potential model, the $S$-factor at low energies is predicted to be dominated by $E$2 transition to the $^{16}$O ground state.
  The $E$1 and $E$2 $S$-factors at $E_{c.m.}=0.3$ MeV are $S_{E1}\approx3$ keV~b and $S_{E2}=150^{+41}_{-17}$ keV~b, respectively.
  The sum of the cascade transition through the excited state of $^{16}$O is $S_{\rm casc}= 18\pm4.5$ keV~b.
  The $\gamma$-ray angular distribution seems to be reproduced with the potential model, whereas the $E$1 $S$-factor at low energies appears to deviate from the experimental value.
  We do not invoke a compensatory large contribution because of the weak coupling feature of the system.

  The derived reaction rates at low temperatures seem to be concordant with CF88, NACRE, KU02, and BU96.
  For astrophysical applications, our reaction rates below $T_9=3$ have been provided in the analytic expression.

\acknowledgments
  The author thanks Professor I.J.~Thompson for his comments on the manuscript of the previous publication.
  He also thanks Y.~Kond\=o for early days of collaboration and Y. Ohnita and Y.~Sakuragi for their hospitality.
  He is grateful to M.~Arnould, A.~Jorissen, K.~Takahashi, and H.~Utsunomiya for their hospitality and encouragement during his stay at Universit\'e Libre de Bruxelles.

\end{document}